\documentclass[structabstract]{aa}  
\usepackage{amsmath}
\usepackage{graphicx}
\usepackage{txfonts}
\usepackage{natbib,twoopt}
\usepackage{color}
\usepackage{multirow}

\begin{document}

        \title{Robust high-contrast companion detection from interferometric observations}
        \titlerunning{Robust high-contrast companion detection from interferometric observations}
        \subtitle{The CANDID algorithm and an application to six binary Cepheids}

        \author{ A.~Gallenne\inst{1},
                                A.~M\'erand\inst{2},
                                P.~Kervella\inst{3, 4}, 
                                J.~D.~Monnier\inst{5},
                                G.~H.~Schaefer\inst{6},
                                F.~Baron\inst{6,7},
                                J.~Breitfelder\inst{3},
                                J.~B.~Le~Bouquin\inst{8},
                                R.~M.~Roettenbacher\inst{5},  
                                W.~Gieren\inst{1,10}, 
                                G.~Pietrzy\'nski\inst{1,9},
                                H.~McAlister\inst{6}, T.~ten~Brummelaar\inst{6}, J.~Sturmann\inst{6}, L.~Sturmann\inst{6}, N.~Turner\inst{6}, S.~Ridgway\inst{11} \and S.~Kraus\inst{12}
                                }
                                
        \authorrunning{A. Gallenne et al.}

\institute{Universidad de Concepci\'on, Departamento de Astronom\'ia, Casilla 160-C, Concepci\'on, Chile
        \and European Southern Observatory, Alonso de C\'ordova 3107, Casilla 19001, Santiago 19, Chile
        \and LESIA, Observatoire de Paris, CNRS UMR 8109, UPMC, Universit\'e Paris Diderot, 5 Place Jules Janssen, F-92195 Meudon, France
        \and Unidad Mixta Internacional Franco-Chilena de Astronom\'ia, CNRS/INSU, France (UMI 3386) and Departamento de Astronom\'ia, Universidad de Chile, Camino El Observatorio 1515, Las Condes, Santiago, Chile
        \and Astronomy Department, University of Michigan, 941 Dennison Bldg, Ann Arbor, MI 48109-1090, USA
        \and The CHARA Array of Georgia State University, Mount Wilson CA 91023, USA
        \and Department of Physics \& Astronomy, Georgia State University, 25 Park Place NE, Atlanta GA 30303-2911
        \and Universit\'e Grenoble Alpes, IPAG, F-38000 Grenoble, France and CNRS, IPAG, F-38000 Grenoble, France
        \and Warsaw University Observatory, Al. Ujazdowskie 4, 00-478, Warsaw, Poland
        \and Millenium Institute of Astrophysics, Santiago, Chile
        \and National Optical Astronomy Observatories, 950 North Cherry Avenue, Tucson AZ 85719, USA
        \and School of Physics, University of Exeter, Stocker Road, Exeter EX4 4QL, UK
        }
  
  \offprints{A. Gallenne} \mail{agallenne@astro-udec.cl}

   \date{Received February 17, 2015; accepted May 4, 2015}

 
  \abstract
   {Long-baseline interferometry is an important technique to spatially resolve binary or multiple systems in close orbits. By combining several telescopes together and spectrally dispersing the light, it is possible to detect faint components around bright stars in a few hours of observations.}
   {We provide a rigorous and detailed method to search for high-contrast companions around stars, determine the detection level, and estimate the dynamic range from interferometric observations.}
   {We developed the code CANDID (Companion Analysis and Non-Detection in Interferometric Data), a set of Python tools that allows us to search systematically for point-source, high-contrast companions and estimate the detection limit using all interferometric observables, i.e., the squared visibilities, closure phases and bispectrum amplitudes. The search procedure is made on a $N \times N$ grid of fit, whose minimum needed resolution is estimated a posteriori. It includes a tool to estimate the detection level of the companion in the number of sigmas. The code CANDID also incorporates a robust method to set a $3\sigma$ detection limit on the flux ratio, which is based on an analytical injection of a fake companion at each point in the grid. Our injection method also allows us to analytically remove a detected component to 1) search for a second companion; and 2) set an unbiased detection limit.}
   {We used CANDID to search for the companions around the binary Cepheids V1334 Cyg, AX Cir, RT Aur, AW Per, SU Cas, and T Vul. First, we showed that our previous discoveries of the components orbiting V1334 Cyg and AX Cir were detected at $>25\sigma$ and $>13\sigma$, respectively. The astrometric positions and flux ratios provided by CANDID for these two stars are in good agreement with our previously published values. The companion around AW Per is detected at more than $15\sigma$ with a flux ratio of $f = 1.22 \pm 0.30$\,\%, and it is located at $\rho = 32.16 \pm 0.29$\,mas and $PA = 67.1 \pm 0.3^\circ$. We made a possible detection of the companion orbiting RT Aur with $f = 0.22 \pm 0.11$\,\%, and at $\rho = 2.10 \pm 0.23$\,mas and $PA = -136 \pm 6^\circ$. It was detected at $3.8\sigma$ using the closure phases only, and so more observations are needed to confirm the detection. No companions were detected around SU Cas and T Vul. We also set the detection limit for possible undetected companions around these stars. We found that there is no companion with a spectral type earlier than B7V, A5V, F0V, B9V, A0V, and B9V orbiting the Cepheids V1334 Cyg, AX Cir, RT Aur, AW Per, SU Cas, and T Vul, respectively. This work also demonstrates the capabilities of the MIRC and PIONIER instruments, which can reach a dynamic range of 1:200, depending on the angular distance of the companion and the $(u, v)$ plane coverage. In the future, we plan to work on improving the sensitivity limits for realistic data through better handling of the correlations.}
   {}

 
 \maketitle

%

\section{Introduction}

Long-baseline interferometry (LBI) enables us to spatially resolve components in close orbits, providing astrometric positions at $<50$\,milli-arcsecond (mas) scale with micro-arcsecond accuracy. When combined with spectroscopic radial velocities, we can obtain model independent estimates of the stellar masses and orbital parallaxes, which are fundamental parameters that help us study stellar properties and evolution. However, so far LBI is limited to bright stars ($H < 7$\,mag) with typical magnitude differences $\Delta H < 6$\,mag. Adaptive optics imaging with single-dish telescopes reaches better contrasts, down to $\Delta H \sim 12$\,mag \citep{Zurlo_2014_10_0}, but the angular resolution is limited to $0.2\arcsec$ at these detection levels. Long-baseline interferometry is therefore a complementary technique in terms of spatial scale by probing the innermost regions.

LBI can detect components down to $\sim 1$\,mas in the infrared, but the main limitation is  sensitivity to high-contrast companions. We roughly know the performances reachable by  current interferometric combiners. So far, the faintest companion detected with LBI has a flux ratio of 0.75\,\% and orbits a Cepheid star of magnitude $H = 3.85$\,mag \citep{Gallenne_2014_01_0}. Objects, such as brown dwarfs and hot giant planets, are still inaccessible because of a lack of sensitivity and accuracy of the instruments. \citet{Absil_2011_11_0} demonstrated a possible dynamic range of 1:500 with the VLTI/PIONIER instrument \citep{Le-Bouquin_2011_11_0}, but this range has not yet been reached. To achieve this detection level, several hours of multitelescope observations are required to obtain as many simultaneous interferometric measurements as possible.

Deriving the sensitivity limit from imaging can be determined directly from the noise level, however, this is not the case for interferometric observations. There are some papers in the literature that discuss detection limits and methods to search for companions from interferometric data \citep{Absil_2011_11_0, Le-Bouquin_2012_05_0}, but these studies have some shortcomings: the searching method is not formalized, the sigma detection is not robust, and they do not take  the bandwidth smearing into account. Therefore, a robust implementation to search for components does not exist thus far. This kind of method is particularly critical to detect faint companions, as they can be at the sensitivity limit of the instrument or even not be a statistically significant detection (i.e., $< 3\sigma$). We therefore created \texttt{CANDID} (Companion Analysis and Non-Detection in Interferometric Data) to address these aspects. This is a suite of \texttt{Python} tools, which contains two main functions: 1) one to perform a systematic search for faint companions (Sect.~\ref{section__detection_level}); and 2) one to estimate the detection limit from long-baseline interferometric observations (Sect.~\ref{section__detection_limit_of_high_contrast_binaries}). This tool is made available to the community\footnote{Available at \url{https://github.com/amerand/CANDID}}.

\texttt{CANDID} is made for detecting high-contrast, point-source companions orbiting a spatially resolved primary star, although it also works with an unresolved primary and contrast $< 50$\,\%. In this paper, we used \texttt{CANDID} to look for companions in binary Cepheids. We present the first main function to search for companions in Sect.~\ref{section__detection_level}, verify and clarify the detection level of our previously detected faint companions, and we report new detections for other Cepheids. In Sect.~\ref{section__detection_limit_of_high_contrast_binaries}, we present the second main function of \texttt{CANDID} and explain our robust method to set detection limits from interferometric data. In
Sect.~\ref{section__aplication}, we then use this formalism to our set of Cepheid observations to derive the detection limits. We finally present our conclusions in Sect.~\ref{section__conclusion}.

\section{Searching for companions}
\label{section__detection_level}

Our grid search is traditionally performed in a three-dimensional space. Specifically, we vary the position of the component ($\Delta \alpha, \Delta \delta$) and the companion/star flux ratio $f$, and then compute the $\chi^2$ for each of these positions and flux ratios (the third dimension being $f$). The weakness of this method is the resolution of the grid, i.e., if the grid is to too coarse, the detection can be missed.

In addition, for high-contrast binaries and/or low accuracy data, a $\chi^2$ map can show fake or nonsignificant detections. Sometimes, some authors show the estimated $\chi^2$ map with the most probable location of the component without any detection level mentioned, only the reduced $\chi^2$ variation is given. This parameter, however, is not optimal for checking the detection level because it depends on the number of degrees of freedom (dof). The large quantity of data required to detect a faint companion using LBI results in a large dof. For instance, we can have the most probable location with the lowest reduced chi-square, $\chi_r^2$, equal to 1.0, and the highest in the whole map equal to 1.1, i.e., with only a variation of 0.1. With this information, however, there is no way to know if the detection is statistically significant or not. In this section, we therefore present a systematic approach to search for components and set the detection level using the $\chi^2$ and the number of dof. We then applied it to the case of binary Cepheids. We  verify our previous detections for the Cepheids \object{V1334~Cyg} and \object{AX~Cir} first \citep{Gallenne_2013_04_0,Gallenne_2014_01_0}, and then report new detections for \object{RT~Aur} and \object{AW~Per}.

\subsection{Detection method}

A more rigorous approach is to perform a grid of fit using a least-squares algorithm, with a starting grid spacing that is guaranteed to find the global minimum. The grid in question is the 2D grid of starting points for the companion position. For each starting position, a multiparameter fit is performed: the companion position and its flux ratio (possibly the stellar diameters) are adjusted. Each position of the grid leads to a local minimum. Ideally, if the starting grid is fine enough, multiple starting points  lead to the same local minima: this guarantees that all the local minima are explored and that the global minimum is indeed the global minimum. Hence, the criteria to decide if the global minimum is global is  a posteriori. In \texttt{CANDID}, we require (a posteriori) that on average, each unique minimum is reached from two starting points of the starting grid. We also provide statistics on the "traveling" distance of the fit (between the starting and end points), compared to the size of the grid. We require that the median travel distance should be less than $\sqrt{2}/2$ of the size of the square grid. The 10 and 90\,\% percentiles are also provided to the user to assess the typical travel distance.

A systematic search using a grid of fit is an iterative process. First, a coarse starting grid is chosen and the fits are performed. \texttt{CANDID} estimates, based on the traveling distances of the fits, how to refine the starting grid. A second series of fits are run, using the finer grid, which might take much longer, but the global minimum can be trusted to be  global (at least within the area searched).

We  searched for components with a maximum distance to the main star of 50\,mas. For a wider range, the loss of coherence caused by spectral smearing of the companion is the main limitation because it degrades the dynamic range. A complete discussion about the search region is presented by \citet{Absil_2011_11_0} and \citet{Le-Bouquin_2012_05_0}. The main limitation is the spectral sampling compared to the relative position of the component, i.e., to avoid a significant smearing we need $R > \rho B_\mathrm{p} /\lambda$, with $R$, $B_\mathrm{p}$ and $\rho$ the spectral resolution, the projected baseline, and the separation, respectively. For PIONIER, with $R = 18$ and a mean projected baseline $B_\mathrm{p} = 100$\,m (used for our observations), we have $\rho < 60$\,mas, and for MIRC with $R = 42$ and $B = 200$\,m, we have $\rho < 70$\,mas. \citet{Zhao_2007_04_0} recommended a more stringent criteria $\rho < R \lambda/(5B)$ to assure uncorrupted data, leading to $\rho \lesssim 15$\,mas for PIONIER and MIRC. However, a companion can still be detected at separations larger than 15\,mas by taking  bandwidth smearing effects into account. We chose 50\,mas as our limiting range to avoid making the grid search too stringent.  In addition, companions located at more than 50\,mas are detected more efficiently using adaptive optics on a single-dish telescope (through imaging or sparce aperture masking).

Each point in the grid is fitted with the following binary model, representing a spatially resolved primary star with a point source component:
\begin{equation}
\label{equation__binaire}
\tilde{V}(u,v) = \frac{V_\star(u,v) + G(\zeta)f\,\tilde{V}_c(u,v)}{1 + f},
\end{equation}
with,
\begin{equation}
\label{equation__UD}
V_\star(u,v) = \frac{2J_\mathrm{1}(x)}{x},
\end{equation}
\begin{equation}
\label{equation__component}
\tilde{V}_c(u,v) = \exp[-2i\pi (u\Delta \alpha + v\Delta \delta)/\lambda],
\end{equation}
\begin{equation}
G(\zeta) = \left| \dfrac{\sin{\zeta}}{\zeta} \right|\quad \mathrm{with}\quad \zeta = \dfrac{\pi (u\Delta \alpha + v\Delta \delta)}{R\lambda} ,
\end{equation}
where $J_\mathrm{1}(x)$ is the first-order Bessel function, $x = \pi \theta_\mathrm{UD}\sqrt{u^2 + v^2}/\lambda$, $(u,v)$ the spatial frequencies, $\theta_\mathrm{UD}$ the uniform disk angular diameter of the primary star, $\lambda$ the wavelength, $f$ the flux ratio between the companion and the primary star, $(\Delta \alpha,\Delta \delta)$ the relative position of the component with respect to the primary, and $R = \lambda / \Delta \lambda$ the spectral resolution. The function $G(\zeta)$ is a corrective term to overcome the effect of bandwidth smearing \citep{Lachaume_2013_11_0}. The fitted parameters are $\theta_\mathrm{UD}, f, \Delta \alpha,$ and $\Delta \delta$. Although the bandwidth smearing can be fit in \texttt{CANDID}, we kept it fixed for this work because we noticed that it cannot be constrained by these data.

The interferometric observables, i.e., the squared visibility $V^2$, the closure phase $CP$, and the bispectrum amplitude $B_\mathrm{amp}$, are then estimated from the squared modulus and the bispectrum in closed triangles as follows:\begin{equation}
V^2 = | \tilde{V} |^2 \quad \mathrm{and} \quad \tilde{B} = \tilde{V}_{12}\tilde{V}_{23}\tilde{V}^*_{31},
\end{equation}
which provides the bispectrum amplitude and the closure phase from the definition $\tilde{B} = B_\mathrm{amp} e^{-iCP}$:
\begin{equation}
B_\mathrm{amp} = |\tilde{B}| \quad \mathrm{and} \quad CP = \mathrm{arg}(\tilde{B}) 
.\end{equation}

The data are then fitted simultaneously using a Levenberg-Marquardt least-squares minimization algorithm with
\begin{equation}
\begin{split}
\chi^2 =  &\sum{ ( CP_\mathrm{o} - CP_\mathrm{m} )^2 / \sigma^2_\mathrm{CP} } + \sum{ ( B_\mathrm{amp,o}-B_\mathrm{amp, m} )^2 / \sigma^2_\mathrm{B_{amp}}} \\
                         &+ \sum{ (V^2_\mathrm{o} - V^2_\mathrm{m} )^2 / \sigma^2_{V^2} },
\end{split}
\end{equation}
where the indexes $o$ and $m$ denote the data and the model, respectively. \texttt{CANDID} can fit all observables or just one, depending on the data available and the user. 
We then divided by the number of degrees of freedom to obtain a map of the chi-square minima, and interpolated these minima on a regular grid.

The grid resolution is critical as it depends on the distance explored from the initial and final positions. This is why we implemented  an estimate of the optimum grid resolution in our code. It is worth mentioning that a denser starting grid is not necessarily better as the fit would not be improved, and we would lose in computation time (\texttt{CANDID} performs a $30 \times 30$ grid with 1\,mas steps in 50\,s using six cores and a data set with 879 degrees of freedom). To make the grid search faster, \texttt{CANDID} was developed for parallel processing on multicore machines.

Once the map of the minima is computed, we can check the variation in the whole map, however this does not tell us about the statistical significance: specifically, whether the most probable location is detected at $1\sigma$ or more. Although it might not be important for low-contrast companions for which the variation is large enough, this is critical for components with a flux ratio $< 5$\,\% with a variation of the minima of a few percent. The number of dof is an important parameter in that context. Assuming that the data follow Gaussian statistics, we implemented in \texttt{CANDID} an estimate of the number of sigma for each point in the grid in order to obtain a $n\sigma$ detection map. The formalism we used is based on the probability $P$ (or confidence interval) with $\nu$ degrees of freedom, as already employed by \citet{Absil_2011_11_0}. The number of sigma demonstrates how our binary model is significant compared to a uniform disk model (i.e., a single star). We used the following formula for the probability:
\begin{equation}
\label{equation__prob}
P (\Delta \alpha ,\Delta \delta) = 1 - CDF_\nu \left(\frac{\nu \chi_\mathrm{UD}^2}{\chi_\mathrm{r, bin}^2(\Delta \alpha ,\Delta \delta)} \right)
,\end{equation}
where $\chi_\mathrm{bin}^2$ and $\chi_\mathrm{UD}^2$ are the minimum chi-square for the binary and the uniform disk model (i.e., fitting eq.~\ref{equation__binaire} and \ref{equation__UD}, respectively), and $CDF$ denotes the $\chi^2$ cumulative probability distribution function with $\nu$ degrees of freedom. We then convert the probability into the number of sigmas, $n\sigma$ (e.g., 99.73\,\% = 3$\sigma$, 99.99\,\% = 4$\sigma$, ... ). To avoid big float numbers, we limited the maximum value to $50\sigma$.

This formalism therefore provides a $\chi_r^2$ map to find the most probable location of a companion, if any, and a $n\sigma$ map giving the detection level at each point in the grid.

\begin{table*}[!ht]
\centering
\caption{Journal of the observations.}
\begin{tabular}{cccccccc} 
\hline
\hline
UT                                               & MJD                                          &       Star    &       $N_\mathrm{bracket}$                            &               Configuration   & $N_{V^2}$       &       $N_{CP}$         & Calibrators  \\
\hline
2012~Jul.~26    &       56135.449               &       V1334~Cyg       &         2       &       S1-S2-E1-E2-W2  &       48      &       42              &       \object{HD~200577}, \object{HD~214200}              \\ 
2012~Sep.~30    &       56201.221               &       V1334~Cyg       &         3       &       S1-S2-E1-E2-W1-W2       &       62      &       68      &       \object{HD~185395}, \object{HD~199956},     \\ 
                                                &                                                               &                               &                 &               &               &               &        \object{HD~218470}, \object{HD~207978}  \\ 
2013~Jul.~14    &       56487.983               &       AX~Cir  &       6       &       D0-H0-G1-I1     &       300     &       200             &       \object{HD~133869}, \object{HD~129462}              \\ 

2012~Jul.~26    &       56134.354               &       T~Vul   &       2       &       S1-S2-E1-E2-W1-W2       &       199     &       260             &       \object{HD~192518}, \object{HD~205852}              \\ 

2012~Sept.~30 & 56200.228               &       T~Vul   &       2       &       S1-S2-E1-E2-W1-W2       &       175     &       210     &       \object{HD~189849}, \object{HD~198692},\\
                                                &                                                               &                               &                 &               &               &               &        \object{HD~207978}\\ 

2012~Sept.~30   &       56200.434 &             SU~Cas  &       4       &       S1-S2-E1-E2-W1  &       259     &       255     &       \object{HD~12216}, \object{HD~19267},                      \\
                                                &                                                               &                               &                 &               &               &               &        \object{HD~34200}\\ 

2012~Oct.~01 &  56201.428               &       AW~Per  &       1               &       S1-S2-E1-E2-W1-W2       &       210     &       280     &       \object{HD~19845}, \object{HD~30825},\\
                                                &                                                               &                               &                 &               &               &               &        \object{HD~35940}\\ 
2012~Oct.~01 &  56201.507               &       RT~Aur  &       1               &       S1-S2-E1-E2-W1-W2       &       105     &       140     &       \object{HD~48682}       \\

\hline
\end{tabular}
\tablefoot{$N_\mathrm{bracket}$: number of data blocks. $N_{V^2}$ and $N_{CP}$: number of squared visibilities and closure phase. Adopted calibrator diameters: HD~200577 = $0.758 \pm 0.052$\,mas, HD~214200 = $0.790 \pm 0.050$\,mas, HD~185395 = $0.726 \pm 0.014$\,mas, HD~199956 = $0.603 \pm 0.043$\,mas, HD~218470 = $0.477 \pm 0.033$\,mas, HD~207978 = $0.571 \pm 0.040$\,mas, HD~133869 = $1.043 \pm 0.015$\,mas, HD~129462 = $0.857 \pm 0.061$\,mas, HD~192518 = $0.418 \pm 0.029$\,mas, HD~205852 = $0.461 \pm 0.032$\,mas, HD~189849 = $0.510 \pm 0.036$\,mas, HD~198692 = $0.660 \pm 0.056$\,mas, HD~207978 = $0.571 \pm 0.040$\,mas, HD~12216 = $0.467 \pm 0.033$\,mas, HD~19267 = $0.586 \pm 0.042$\,mas, HD~34200 = $0.652 \pm 0.046$\,mas, HD~48682 = $0.616 \pm 0.043$\,mas, HD~19845 = $0.788 \pm 0.056$\,mas, HD~30825 = $0.564 \pm 0.040$\,mas, HD~35940 = $0.615 \pm 0.044$\,mas}
\label{table__journal}
\end{table*}

\begin{figure*}[!t]
\centering
\resizebox{\hsize}{!}{\includegraphics[width = \linewidth]{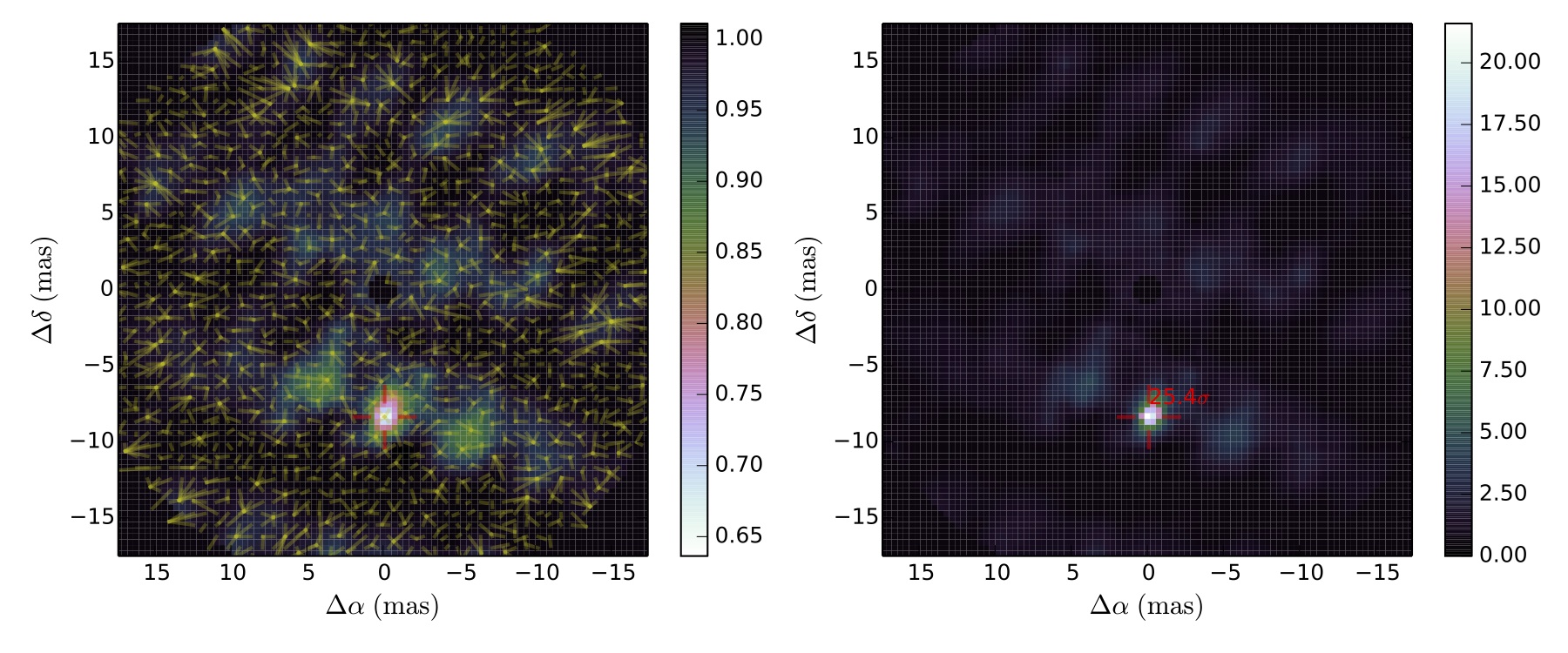}}
\caption{$\chi_r^2$ map of the local minima (left) and detection level map (right) of V1334~Cyg for the observations on 2012-10-01. The yellow lines represent the convergence from the starting points to the final fitted position. The maps were reinterpolated in a regular grid for clarity. The axis limit was chosen according to the location of the companion.}\ 
\label{image__v1334}
\end{figure*}
\begin{figure*}[!t]
\centering
\resizebox{\hsize}{!}{\includegraphics[width = \linewidth]{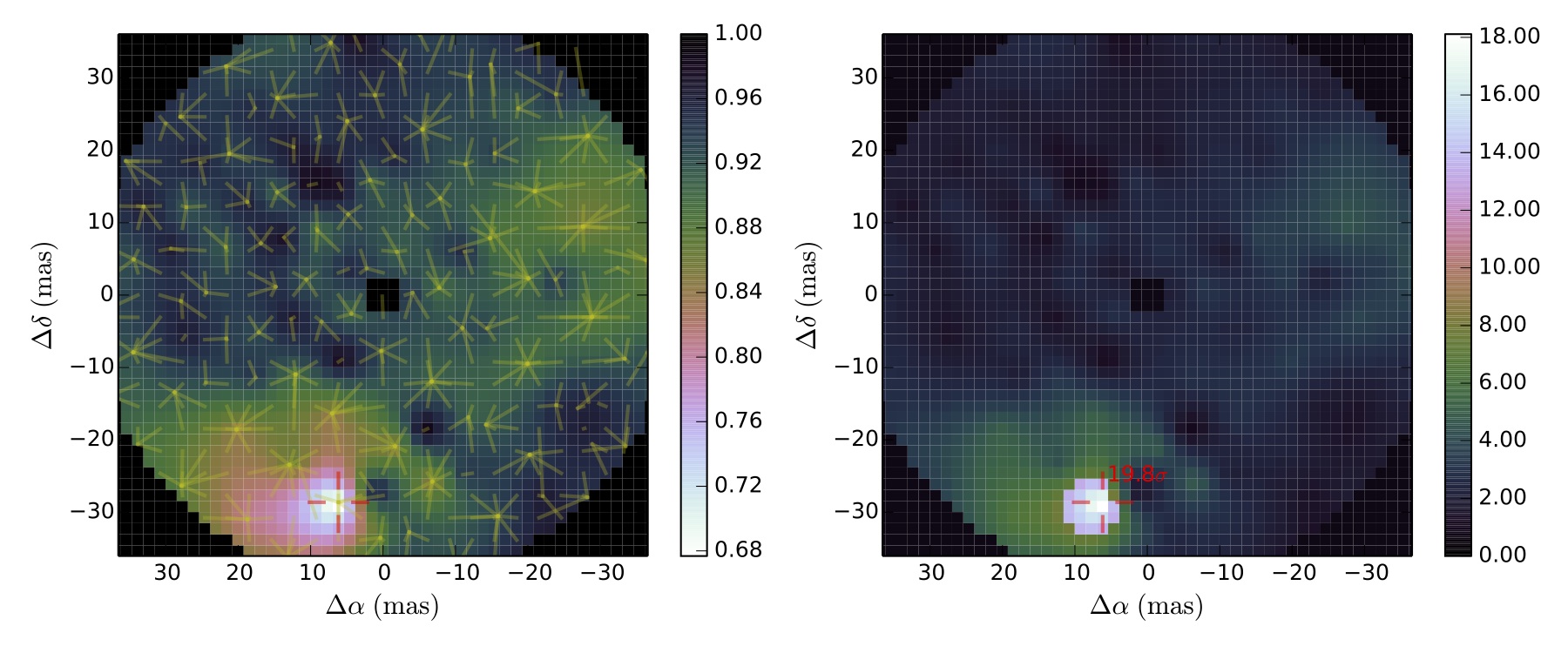}}
\caption{Same as Fig.~\ref{image__v1334}, except for AX~Cir for the observations on 2013-07-14.}
\label{image__axcir}
\end{figure*}

\subsection{Previously published detections}

To validate our method, we computed the maps of the $\chi^2$ minima and estimated the detection level for our previously detected companions around the Cepheids \object{V1334~Cyg} and \object{AX~Cir} \citep[][observed with the instruments CHARA/MIRC and VLTI/PIONIER, respectively]{Gallenne_2013_04_0,Gallenne_2014_01_0}. The journal of these previous observations are reported in Table~\ref{table__journal}. The maps are presented in Fig.~\ref{image__v1334} and \ref{image__axcir}, for which all observables were fitted, except for PIONIER where only $CP$ and $V^2$ were used because there is no good estimator of $B_\mathrm{amp}$ so far. The central part has been hidden to improve the clarity of the intensity map level, which can be biased by the primary star. The companion orbiting V1334~Cyg is detected at more than $25\sigma$, and at more than $13\sigma$ for AX~Cir. We summarized the detection levels for these two stars in Table~\ref{table__detection_level}, including fitting only the closure phase signal. The CP is more sensitive to faint off-axis companions and is also less affected by instrumental and atmospheric perturbations than the other observables (i.e., $V^2$ and $B_\mathrm{amp}$). Fitting all of the observables can improve the detection level because we add more information, but it can also affect the results, depending on the magnitude of the biases. We  notice from Table~\ref{table__detection_level} that including $V^2$ and $B_\mathrm{amp}$ degrades the detection level for V1334~Cyg, although it is still  significant, and including these observables improves the detection for AX~Cir. A possible explanation that addresses why the detection level decreases sometimes when we add the $V^2$ is that we add correlated noise, which is not consistent with our hypothesis of uncorrelated noise. When only the CP is used, the angular diameter of the primary is first determined by fitting only a uniform disk model to the square visibilities, and then kept fixed during the grid search.

The resulting fitted parameters (i.e., the astrometric position, flux ratio, and the angular diameter) are in good agreement with the values determined in \citet{Gallenne_2013_04_0, Gallenne_2014_01_0}.

\subsection{New detections of the companions around the Cepheids RT~Aur and AW~Per}
\label{subsection__new_detection}

Our interferometric program on Galactic binary Cepheids, which started two years ago, is promising to directly measure the dynamical masses. We have obtained several observing nights in both hemispheres with the multitelescope combiners CHARA/MIRC and VLTI/PIONIER to detect the close companions of a few Cepheids \citep{Gallenne_2014_01_0,Gallenne_2013_04_0,Gallenne_2013_02_0}. Here we report new detections for the Cepheids AW~Per and RT~Aur.

The observations were performed in 2012 using the Michigan InfraRed Combiner (MIRC) installed at the CHARA Array \citep{ten-Brummelaar_2005_07_0}, located on Mount Wilson, California. The CHARA Array consists of six 1\,m aperture telescopes in a Y-shaped configuration (two telescopes on each branch), oriented to the east (E1, E2), west (W1,W2) and south (S1, S2), providing good coverage of the $(u, v)$ plane. The baselines range from 34\,m to 331\,m, providing high angular resolution down to 0.5\,mas at $H$. The MIRC instrument \citep{Monnier_2004_10_0,Monnier_2010_07_0} is an image-plane combiner, which enables us to combine the light coming from all six telescopes in $H$ or $K$. MIRC also offers three spectral resolutions ($R = 42, 150$ and 400), which provide 15 visibility and 20 closure phase measurements across a range of spectral channels.

\begin{figure*}[!t]
\centering
\resizebox{\hsize}{!}{\includegraphics[width = \linewidth]{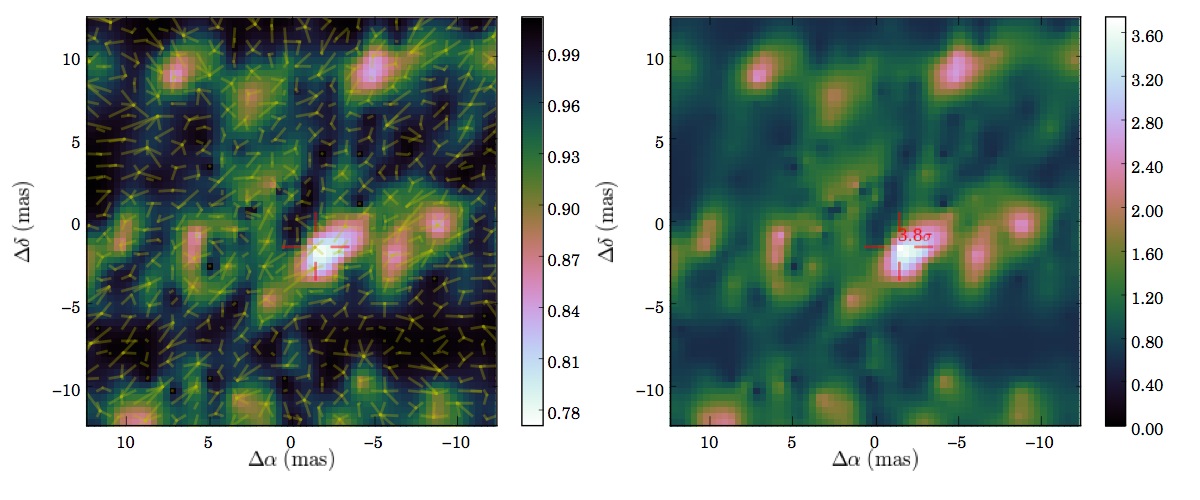}}
\caption{Same as Fig.~\ref{image__v1334}, except for RT~Aur using only the closure phase.}
\label{image__rtaur}
\end{figure*}
\begin{figure*}[!t]
\centering
\resizebox{\hsize}{!}{\includegraphics[width = \linewidth]{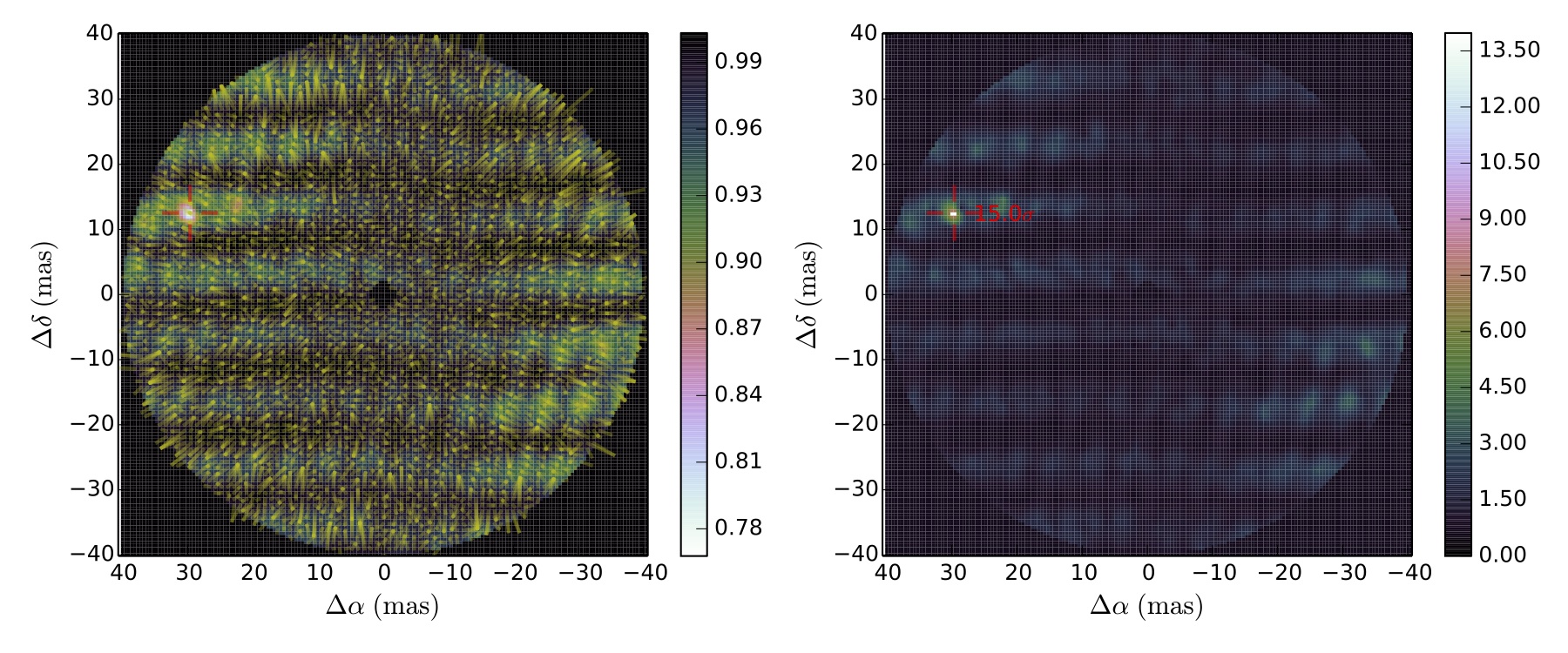}}
\caption{Same as Fig.~\ref{image__v1334}, except for AW~Per.}
\label{image__awper}
\end{figure*}

\begin{table}[!h]
\centering
\caption{Summary of the detection level for our previous published data and new detections.}
\begin{tabular}{ccccc} 
\hline
\hline
Star                    &               $CP$ only               &       ALL \tablefootmark{a} & dof         & Instrument\\
                                        &               ($n\sigma$)     &       ($n\sigma$) &                       \\
\hline          
V1334~Cyg       &                       42.1                            &    25.4                 & 1583  &       MIRC                    \\ 
AX~Cir                  &                       13.6                            &    19.8         &       1499                    & PIONIER       \\ 
RT~Aur                                  &         3.8                           &  2.3                     &    879               & MIRC  \\
AW~Per                                  &         19.6                          &         15.0                    &       1757    & MIRC  \\ 
\hline
\end{tabular}
\tablefoot{\tablefoottext{a}{All means $CP$ + $V^2$ + $B_\mathrm{amp}$, and $B_\mathrm{amp}$ is only used with MIRC data.}}
\label{table__detection_level}
\end{table}

We observed the short-period Cepheids \object{RT~Aur} (HD~45412, $P_\mathrm{puls} = 3.73$\,d) and \object{AW~Per} (HD~30282, $P_\mathrm{puls} = 6.46$\,d) with all six telescopes. We used the $H$-band filter with the lowest spectral resolution in which the light is split into eight spectral channels. Table~\ref{table__journal} lists the journal of our observations. We followed a standard observing procedure in which we monitored the interferometric transfer function by observing a calibrator before and/or after the Cepheids. The calibrators, listed in Table~\ref{table__journal}, were selected using the \textit{SearchCal}\footnote{Available at http://www.jmmc.fr/searchcal.} software \citep{Bonneau_2006_09_0,Bonneau_2011_11_0} provided by the Jean-Marie Mariotti Center\footnote{http://www.jmmc.fr}.

We reduced the data using the standard MIRC pipeline \citep{Monnier_2007_07_0}, which consists of computing the squared visibilities and triple products for each baseline and spectral channel, and correcting for photon and readout noise. Squared visibilities are estimated using Fourier transforms, while the triple products are evaluated from the amplitudes and phases between three baselines forming a closed triangle.

We used \texttt{CANDID} to search for a component within $\pm 50$\,mas. For RT~Aur, we might have detected a companion at $3.8\sigma$ using only the closure phases, while we only have  a $2.3\sigma$ detection using all observables. The possible companion is detected at $\rho = 2.1$\,mas and $PA = -136^\circ$ with a flux ratio $f = 0.22$\,\%. The component orbiting AW~Per is detected at $> 15\sigma$ at $\rho = 32$\,mas and $PA = 67^\circ$ with a flux ratio $f = 1.22$\,\%. The grids of fit and detection level maps are shown in Fig~\ref{image__rtaur} and \ref{image__awper}, and the final fitted parameters are listed in Table~\ref{table__fitted_parameters}. For the uncertainties, we used the conservative formalism of \citet{Boffin_2014_04_0} for all the fitted parameters, i.e.,
\begin{equation}
\sigma^2_\mathrm{X} = N_\mathrm{sp}\sigma_\mathrm{stat}^2 + 0.0001X^2
,\end{equation}
where $N_\mathrm{sp}$ is the number of spectral channels and $X$ denotes the fitted parameters (i.e., $\Delta x, \Delta y, f$ and $\theta_\mathrm{UD}$). The first term takes into account that the spectral channels are almost perfectly correlated, and the second term comes from the fact that the wavelength calibration is only precise at a 1\,\% level. The parameter $\sigma_\mathrm{stat}$ is the statistical error from the bootstrapping technique (bootstrap on the calibrated data with replacement) using 10~000 bootstrap samples (also included in \texttt{CANDID}). We then took from the distributions the median and the maximum value between the 16\,\% and 84\,\% percentiles as the uncertainty (although the distributions were roughly symmetrical).

There are additional significant peaks in the RT~Aur maps, with a detection level $> 3\sigma$, however, these are spuriously produced by the ($u,v$) coverage and the presence of the companion. It is worth mentioning that RT~Aur was only observed for  one hour (one sequence), and we need more data to confirm the presence of the companion. A more complete discussion about the detected companions is presented in Sect.~\ref{section__aplication}.

\begin{table*}[]
\centering
\caption{Final best-fit parameters.}
\begin{tabular}{ccccc} 
\hline
\hline
                                                                                                                &       RT~Aur                                                  &       AW~Per                                                         &  SU~Cas                               &       T~Vul           \\
                                                                                                                &               $\phi = 0.32$                                 &       $\phi = 0.52$                   &       $\phi = 0.77$                 &       $\phi = 0.27$ / $\phi = 0.12$ \\
\hline
$\theta_\mathrm{UD}$ (mas)                              &       $0.699 \pm 0.011$           &      $0.627 \pm 0.018$               &   $0.609 \pm 0.043$     & $0.608 \pm 0.013 / 0.635 \pm 0.018$       \\
$f$ (\%)                                                                                        &       $0.22 \pm 0.11$                               &       $1.22 \pm 0.30$         &       --      &       --\\
$\Delta \alpha$ (mas)                                           &       $-1.458 \pm 0.238$     &        $29.624 \pm 0.305$      &   --           &      --\\
$\Delta \delta$ (mas)                                    &      $-1.506 \pm 0.224$                  &       $12.523 \pm 0.147$      &       --   &  --\\
\hline
\end{tabular}
\tablefoot{$\theta_\mathrm{UD}$: uniform disk angular diameter, respectively. $f$, $\Delta x$, $\Delta y$: flux ratio and position of the companion.
}
\label{table__fitted_parameters}
\end{table*}

\begin{figure*}[!t]
\centering
\resizebox{\hsize}{!}{\includegraphics[width = \linewidth]{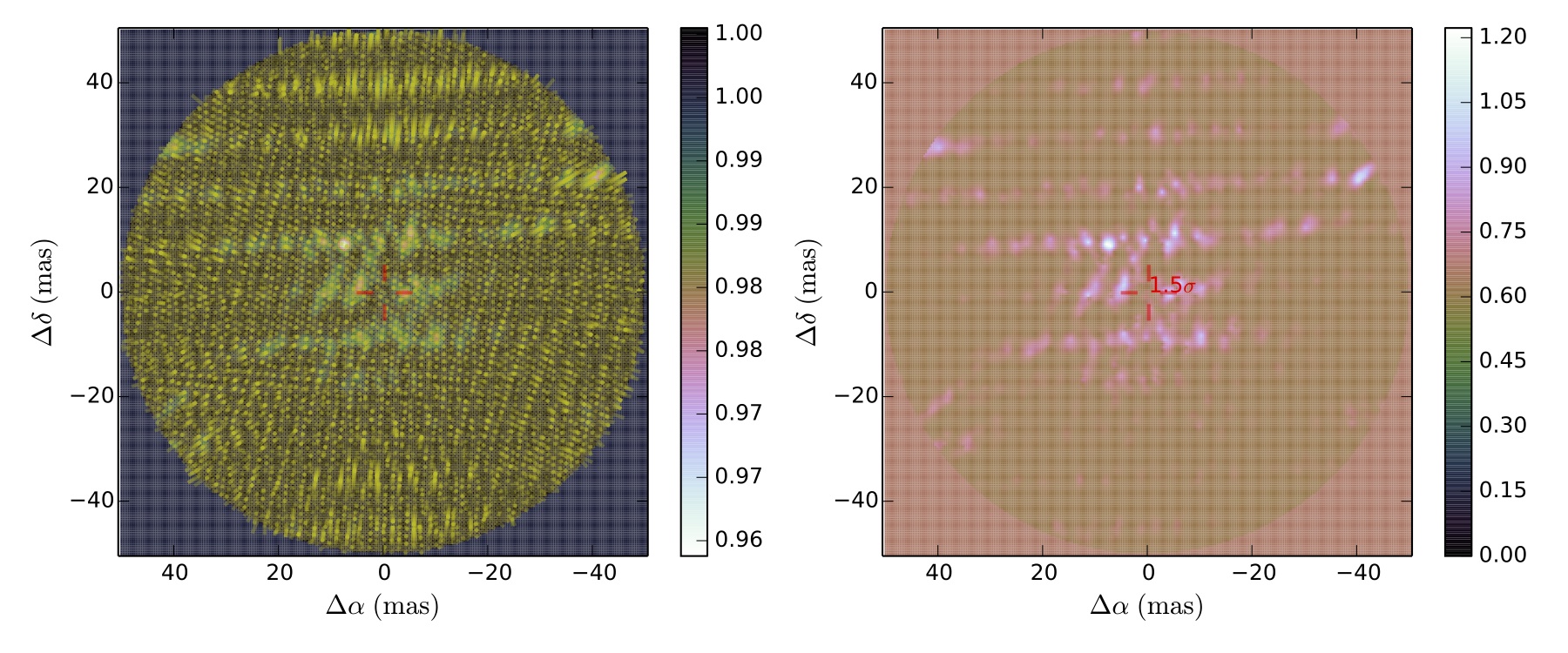}}
\caption{Same as Fig.\ref{image__rtaur}, except we analytically removed the companion.}
\label{image__rtaurnocomp}
\end{figure*}

\subsection{Undetected companions}

We also observed with MIRC the Cepheids SU~Cas (HD~17463, $P = 1.95$\,day) and T~Vul (HD~198726, $P = 4.44$\,day). We used five and six telescope configurations, using the same instrument setup and calibration procedures as explained in the previous section. Table~\ref{table__journal} lists the journal of these observations.

We did not detect any significant companions around these stars, either using all of the observables or only the $CP$, i.e., no more than $1.9\sigma$ and $2.9\sigma$ for SU~Cas and T~Vul, respectively. However, we were able to measure their angular diameters, which are listed in Table~\ref{table__fitted_parameters}. A more detailed discussion of these binary systems is presented in Sect.~\ref{section__aplication}.

\section{Detection limit of high-contrast binaries}
\label{section__detection_limit_of_high_contrast_binaries}

It is important to be able to check the dynamic range that can be reached with a given set of data and any interferometric combiner. \citet{Absil_2011_11_0} already presented a method to set detection limits for the VLTI/PIONIER instrument, but it has some shortcomings, that is why we propose a more robust formalism.

\paragraph{Absil's method:} Their method is based on comparing a uniform disk model with a binary model for each position ($\Delta \alpha, \Delta \delta$) in the grid. They then checked whether the probability of the binary model is consistent with the data using,
\begin{equation}
\label{equation__prob_absil}
P (\Delta \alpha ,\Delta \delta) = 1 - CDF_\nu \left(\frac{\nu \chi_\mathrm{r, bin}^2(\Delta \alpha ,\Delta \delta)}{\chi_\mathrm{UD}^2} \right)
.\end{equation}
We notice that this equation has a different ratio in the $CDF$ than Eq.~\ref{equation__prob}. This is because we assumed that the binary model is the true model, while \citet{Absil_2011_11_0} assumed the uniform disk as the true model. In theory, both equations should lead to the same results, however, as we  see in subsequent sections, their method is more sensitive to biased data and can sometimes lead to under- or overestimated detection limits.

\paragraph{Our method:} We suggest an alternative method, which is based on the injection of a companion into the data at each astrometric position with different flux ratios. As we inject a companion, we therefore know that the binary model should be the true model, and we can use Eq.~\ref{equation__prob} to obtain the probability of the binary model to be the true model. We introduced this method because we think it is more robust, as we  demonstrate in the next section.

In this section, we first introduce approximate formulae for high-contrast companions. We then explain how we inject an additional component into the data and derive the detection limits.

\subsection{High-contrast approximation}

The complex visibility for a binary system composed of a resolved primary star and an unresolved component is given by Eq.~\ref{equation__binaire}. The squared visibility is given by,

\begin{equation}
\label{equation__vis_approx}
\begin{split}
V^2 = & | \tilde{V}.\tilde{V}^* | = \dfrac{1}{1 + f^2} | ( V_\star + G(\zeta) f e^\mathrm{i\varphi} ) ( V_\star + G(\zeta) f e^\mathrm{-i\varphi} ) | \\
                = & \left| \dfrac{V^2_\star + 2 G(\zeta) f V_\star \cos \varphi + G(\zeta)^2 f^2}{(1 + f)^2} \right|, 
\end{split}
\end{equation}
where we kept the $f^2$ term to avoid having an (small) offset (see next section). For a high-contrast companion, i.e., for $f << 1$, we can approximate the bispectrum at the first order in $f$ as,

\begin{equation}
\begin{split}
\tilde{B} = & \dfrac{( V_1 + G(\zeta_1) f e^\mathrm{i\varphi_1} ) ( V_2 + G(\zeta_2) f e^\mathrm{i\varphi_2} ) ( V_3 + G(\zeta_3) f e^\mathrm{-i\varphi_3} )}{(1 + f)^3}  \\
                         \sim& \dfrac{ \tilde{B}_\star}{(1+f)^3} \left[ 1 + f \left( \dfrac{G(\zeta_1) e^{-i\varphi_1}}{\tilde{V}_{1\star}} +  \dfrac{G(\zeta_2) e^{-i\varphi_2}}{\tilde{V}_{2\star}} +  \dfrac{G(\zeta_3) e^{i\varphi_3}}{\tilde{V}_{3\star}^*} \right) \right]\\
                        \sim& \dfrac{ \tilde{B}_\star}{(1+f)^3}\, .Z
\end{split}
,\end{equation}
where $\tilde{B}_\star = \tilde{V}_{1\star}\tilde{V}_{2\star}\tilde{V}^*_{3\star}$.

The bispectrum amplitude and the closure phase are then estimated as,
\begin{eqnarray}
\label{equation__bispectrum_approx1}
B_\mathrm{amp} &=& \dfrac{ 1}{(1+f)^3}\ |\tilde{B}_\star| \ |Z|\\
\label{equation__bispectrum_approx2}
CP &=& \mathrm{arg}(\tilde{B}_\star) - \mathrm{arg}(Z)
.\end{eqnarray}

\subsection{Adding/removing a component}


If no companion is detected in the data, we can assume that the measured values only represent  the primary star, i.e., a uniform disk, and we can substitute the index "$\star$" in the previous equations by the index "$\mathrm{obs}$" (note that the function $Z$ also depends on the primary star visibility). It is now simple, from Eq.~\ref{equation__vis_approx}, \ref{equation__bispectrum_approx1} and \ref{equation__bispectrum_approx2}, to inject a companion to the observed data, which corresponds to the following equations for the following new observables:
\begin{eqnarray}
\label{equation__approx1}
V^2 &=& \dfrac{V^2_\mathrm{obs} + 2 G(\zeta) f V_\mathrm{obs} \cos \varphi + G(\zeta)^2 f^2}{(1 + f)^2},\\
\label{equation__approx2}
B_\mathrm{amp} &=& \dfrac{ 1}{(1+f)^3} \ B_\mathrm{amp,obs}\ |Z|,\\
CP &=& CP_\mathrm{obs} - \mathrm{arg}(Z)
\label{equation__approx3}
.\end{eqnarray}

The original "oifits" files provide all the necessary spatial and spectral information to reconstruct the individual phases and visibilities. We also take  the coherence loss effect caused by the spectral smearing of the companion into account.

Inversely, if a component is detected, we can use this formula to analytically remove the companion (with a negative flux ratio), and then check for another possible fainter component or estimate the detection limit to rule out any other companions. This step is critical to obtain unbiased detection limits. As an example in Fig.~\ref{image__rtaurnocomp} , we analytically removed the detected component orbiting RT~Aur, and  notice that there is no significant detection of a third component and that the other $\chi^2$ minima are not additional companions.

\begin{figure}[!ht]
\centering
\resizebox{\hsize}{!}{\includegraphics{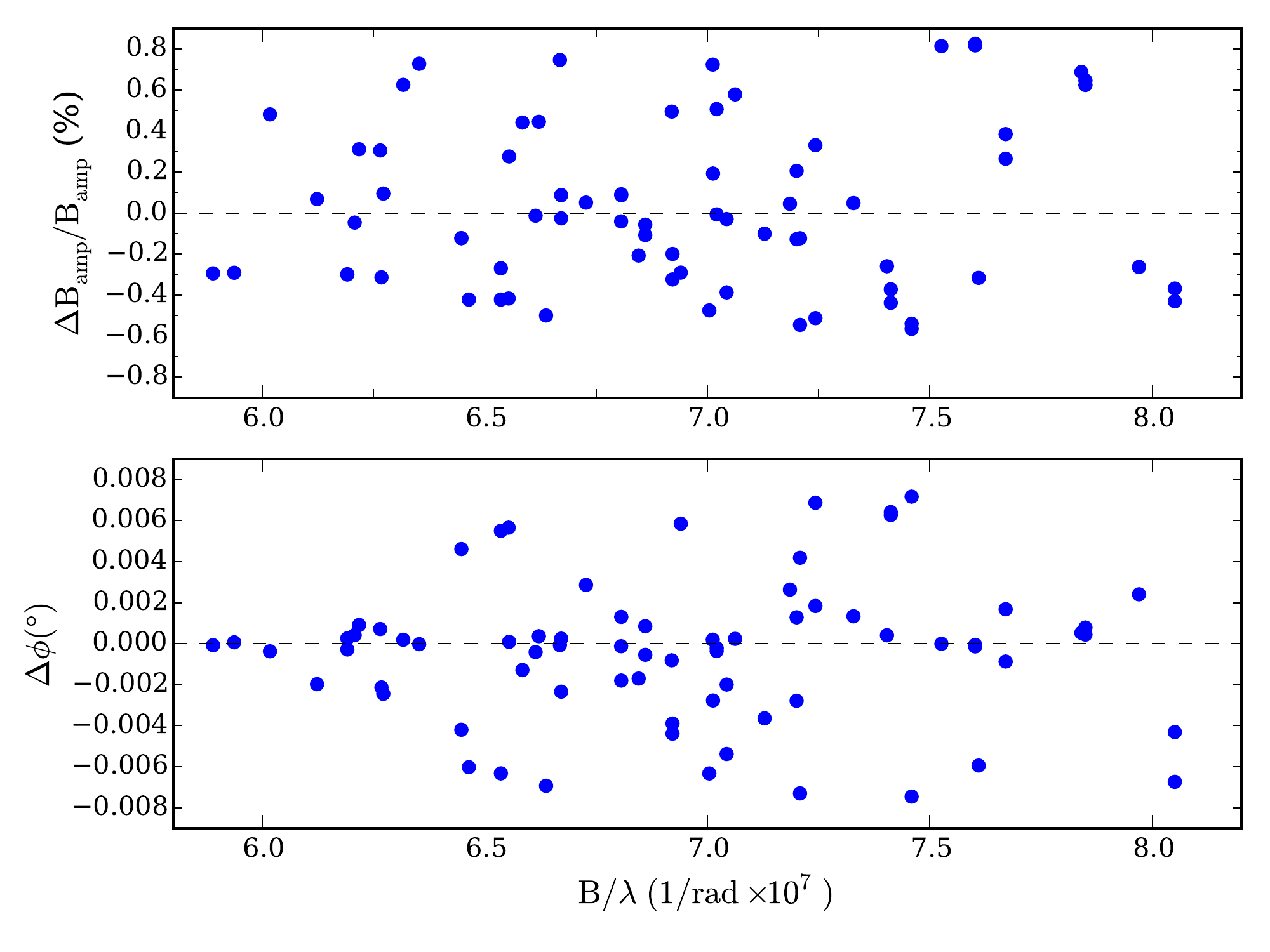}
}
\caption{Difference between the true binary model and the approximation model. For the bispectrum, the abscissa is the maximum of the three spatial frequencies.}
\label{image__test}
\end{figure}

To check our approximation, we created a model of a single star with a uniform disk of 1\,mas (3h of PIONIER observations with a point every 30\,min), for which we added a companion at a position $\Delta \alpha = \Delta \delta = 50$\,mas with a flux ratio $f = 5$\,\%. We then compare this model with a true binary model using the same parameters (Eq.~\ref{equation__binaire}). The difference between the approximation and the true model is shown in Fig.~\ref{image__test}. We found for the amplitude of the bispectrum a standard deviation of the relative error $< 0.3$\,\%, and a difference $< 0.005^\circ$ for the closure phase, which is lower than the achievable interferometric accuracy. Closer companions with the same or higher contrast (i.e., $<5$\,\%) lead to smaller errors. 

We  kept the $f^2$ term in the squared visibility, otherwise it gives an offset in the relative error of about 0.3\,\% for $f = 5$\,\%, and decreases with decreasing flux ratios. Although this value is negligible compared to the current possible data accuracy ($\Delta V^2 / V^2 = \Delta B_\mathrm{amp} / B_\mathrm{amp} \sim 2$\,\%, $\Delta CP \sim 0.5^\circ$), the offset is larger for lower contrasts ($\sim 1.2$\,\% with $f = 10$\,\%). As no approximation is made for $V^2$, the formula is valid for any flux ratio. We therefore also checked the bispectrum approximation for brighter components, and  found that we can use it up to a flux ratio of 50\,\% only using $CP$ and $V^2$. We listed in Table~\ref{table__approx_validity_domain} the relative error for different flux ratios and various positions ($1 < \rho < 50$\,mas and $0 < PA < 2\pi$). We see that \texttt{CANDID} can also be used for low-contrast companions.

\begin{table}[!h]
\centering
\caption{Validity domain of our approximation.}
\begin{tabular}{ccc} 
\hline
\hline
$f$             &       $\Delta CP$ & $\Delta B_\mathrm{amp} / B_\mathrm{amp}$         \\
(\%)                            &       ($^\circ$)    & (\%)                    \\
\hline          
5               &    $< 0.01$           &       $< 0.5$         \\ 
10      &    $< 0.02$           &       $< 1.9$ \\ 
50      &       $< 0.25$                    &    $< 67$         \\
75      &    $< 0.7$            &       $> 100$ \\ 
100     &    $< 3.5$                            &       $> 100$         \\ 
\hline
\end{tabular}
\label{table__approx_validity_domain}
\end{table}

\subsection{Estimating the detection limit}

We implemented in \texttt{CANDID} a tool to estimate the detection limit for a given set of interferometric data. The method is to compare the $\chi^2$ obtained for a model without a companion and the $\chi^2$ obtained for a model with an injected companion. The method does not work, a priori, with an already detected companion, first because the previous approximation might not be valid and also because the resulting detection limit would be biased by a systematic offset related to the flux ratio between the components. This means that any detected component has to be analytically removed first. 

The method is based on an $N\times N$ grid with a range of $\pm 50$\,mas, for which the minimum $N$ depends on the optimum resolution estimated from the $\chi^2$ map (see Sect.~\ref{section__detection_level}). At each point in the grid, we inject a companion with various flux ratios and we compute the $\chi_r^2$. As we know that the true model is the binary model (because we injected a companion), we used our previous equation, Eq.~\ref{equation__prob}, to estimate the number of sigmas for each flux ratio. We then interpolated the flux ratio values at $3\sigma$, which we set as the significance level. This means that lower flux ratios are not   detected significantly. Doing this for all points in the grid, we then have a $3\sigma$ detection limit map for the flux ratio. To have a quantitative estimate of the sensitivity limit with respect to the separation, we estimated a radial profile, $f_{3\sigma}(r)$, using the 90\,\% completeness level (i.e., 90\,\% of all possible positions) from the cumulated histogram in rings for all azimuths.  This tool also includes parallel processing to make the calculation faster.

In theory, for uncorrelated data with Gaussian statistics, our method should lead to the same results as \citet{Absil_2011_11_0} in terms of the detection limit. However, real data are often biased by different sources (atmospheric turbulence, mechanical vibrations, ...). We performed two tests to compare both methods, one including uncorrelated Gaussian noise and another with a noise model, i.e., correlated non-Gaussian noise. We used all observables and three data sets for each test. The \texttt{Aspro2} software\footnote{Available at \url{http://www.jmmc.fr/aspro_page.htm}} were used to create these synthetic data sets, however, it does not have implemented the bandwidth smearing effect. We therefore did not take it into account in \texttt{CANDID} (this would not change the conclusion of our test).

\begin{figure}[]
\centering
\resizebox{\hsize}{!}{\includegraphics{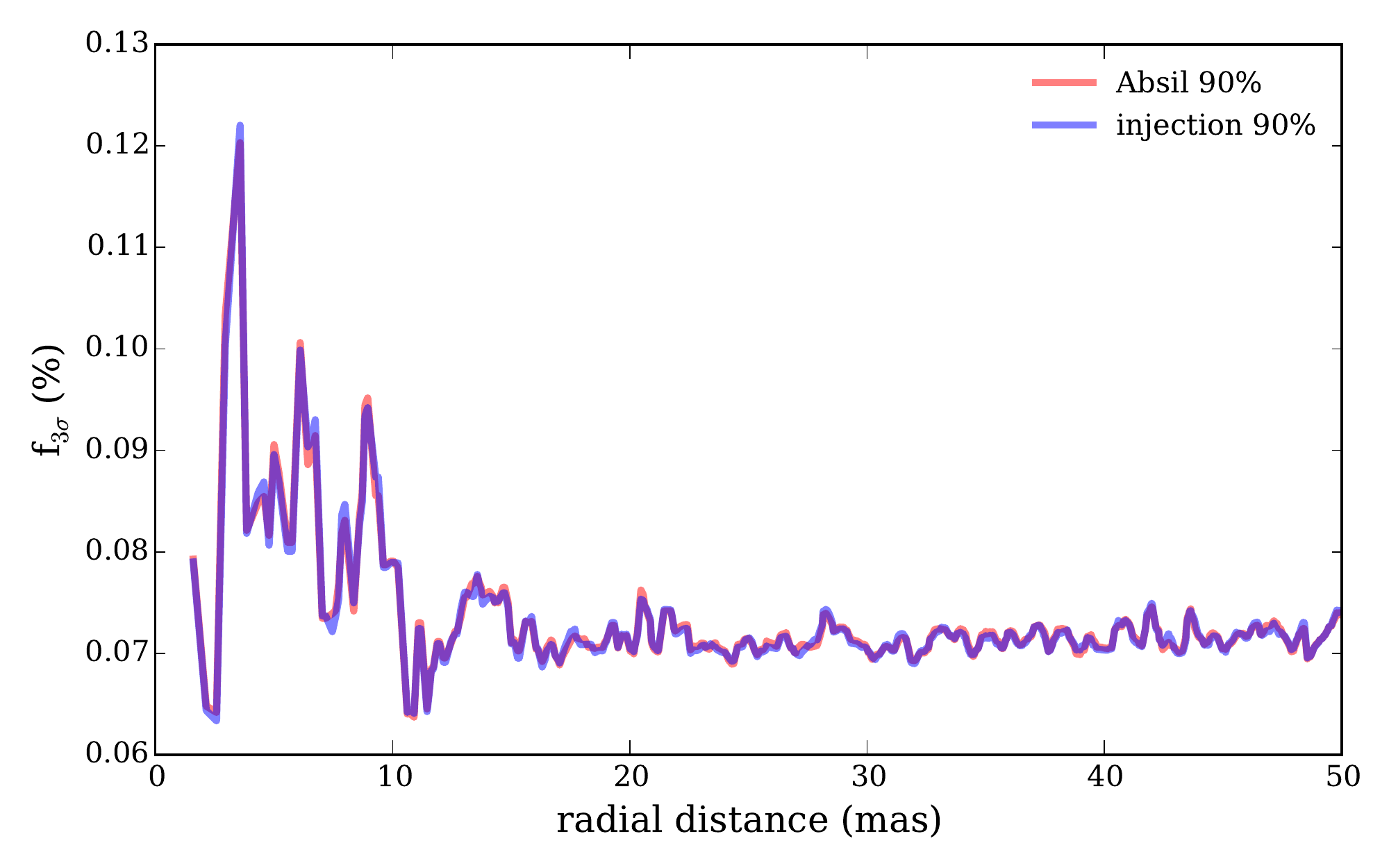}}
\caption{Comparison between the \texttt{CANDID} detection limit tool and the formalism of \citet{Absil_2011_11_0} for a uniform disk model with Gaussian noise.}
\label{image__comparison_ud}
\end{figure}

\paragraph{First test:} We created the first "ideal" synthetic data set (i.e, without noise) representing a uniform disk of 1\,mas (3h observation with three spectral channels with the PIONIER instrument). We then added uncorrelated Gaussian noise and estimated the detection limit using the Absil's method and our formalism. As expected, we see in Fig.~\ref{image__comparison_ud} that both methods provide the same results. We then created a second data set by adding  a component at $\Delta \alpha = \Delta \delta = 5$\,mas with $f = 2$\,\% as a source of bias to the previous model. The companion creates a departure from the ideal measurements (like noise) but in such a way that the observations at different spatial frequencies have a correlated departure (hence, correlated noise). We then estimated the detection limits for this second data set, and although the trend between the two methods are similar, there is an offset of $\sim 2$\,\% because of the presence of the faint companion. Finally, we analytically removed the companion to get our third data set. Comparing the three data sets, i.e., estimating the total variation of the detection limit for each radius, we noticed that both methods vary in a similar way. This was expected as all of the noise sources inserted have Gaussian statistics. However, it is worth mentioning that the scatter tends to be a bit larger for the Absil's method for increasing flux ratios and wider separations of the companion (due to the correlated noise introduced by the component).

\begin{figure}[]
\centering
\resizebox{\hsize}{!}{\includegraphics{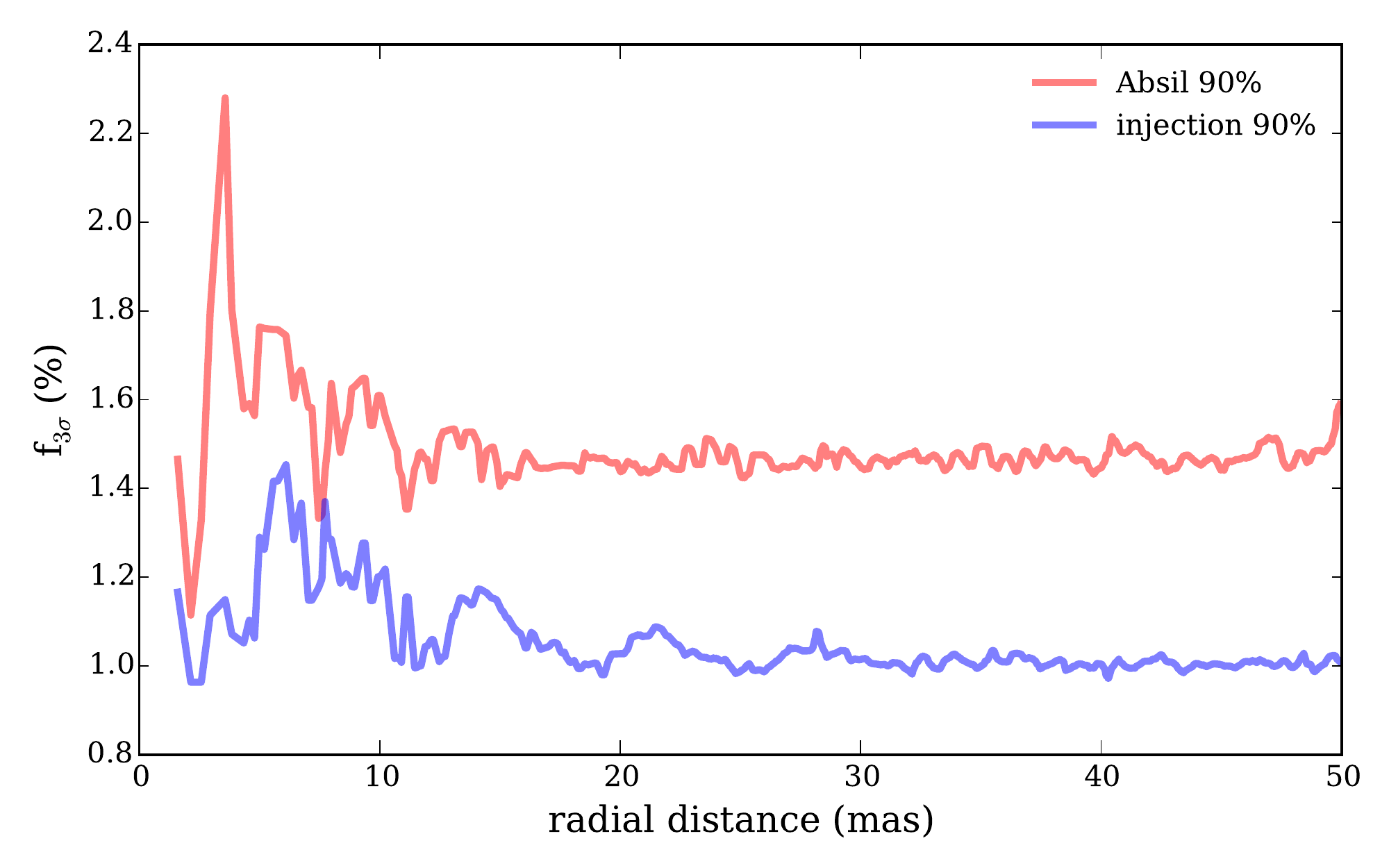}}
\caption{Comparison between the \texttt{CANDID} detection limit tool and the formalism of \citet{Absil_2011_11_0} for a uniform disk model with a noise model, biased by the presence of a faint companion.}
\label{image__comparison_bin}
\end{figure}

\begin{figure}[]
\centering
\resizebox{\hsize}{!}{\includegraphics{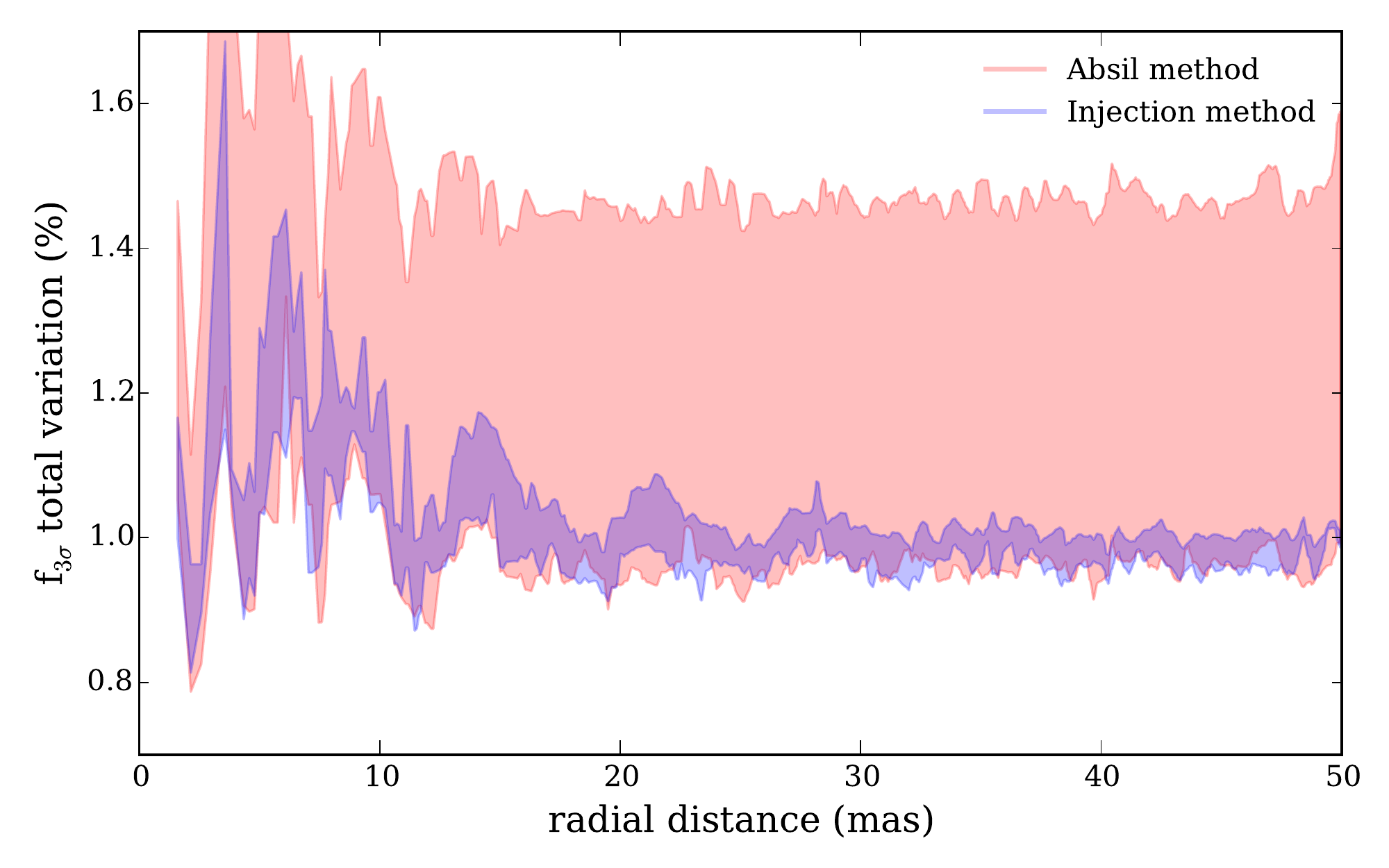}}
\caption{Total variation (minimum and maximum values) between the three synthetic data sets with a noise model for both methods: a single star, a single star + an unresolved component as source of bias, and the previous model with the component analytically removed. The variation has been normalized to unity.}
\label{image__comparison_removed}
\end{figure}

\paragraph{Second test:} We created two new data sets similar to the first test, i.e., one for a uniform disk of 1\,mas and another for a binary system with $\Delta \alpha = \Delta \delta = 5$\,mas and $f = 2$\,\% (with average atmospheric conditions, which is an option for the noise model). The only difference is that the noise is no longer Gaussian; it is represented by a more complicated model that includes the instrument response, atmospheric turbulence, photon, and detector noise\footnote{Details are explained in  \url{http://www.jmmc.fr/doc/approved/JMMC-MEM-2800-0001.pdf}}. As shown in Fig.~\ref{image__comparison_bin}, the absil's method is more sensitive to the presence of the companion, which in this case gives a higher sensitivity limit. The third data set is created by removing the component using our approximation. We then compared the variation of the detection limit between all of the data sets, as described previously, i.e., the minimum and maximum value between the three synthetic data sets at each separation. Fig.~\ref{image__comparison_removed} shows that the scatter from the Absil's method is larger. We performed additional tests with lower flux ratios and at different positions, and the scatter from the Absil's method is larger than our formalism most of the time. We therefore conclude that our method is more robust to biased data when estimating detection limits.

In our tests, the limits derived from the injection method are lower than when we use  Absil's method, but this is not always the case for real data. The injection method occasionally results in higher sensitivity limits, depending on the magnitude of the biases. We find that Absil's method may under- or overestimate the detection limits depending on the data set. However, in some cases when the data are not affected by significant biases, both methods give similar results.


\begin{figure}[]
\centering
\resizebox{\hsize}{!}{\includegraphics{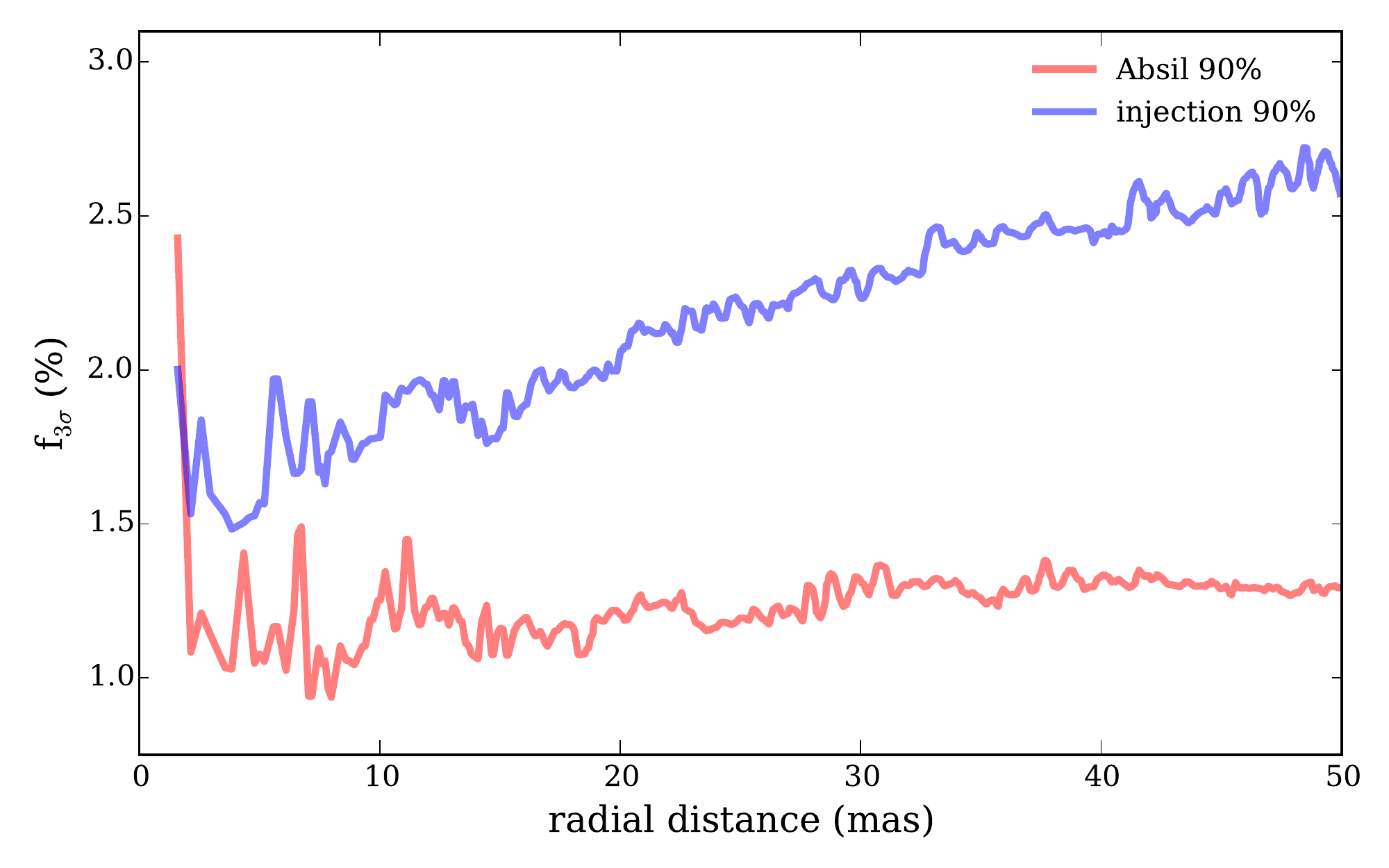}}
\caption{Flux ratio detection limit at $3\sigma$ for a second companion around  V1334~Cyg (for 2012 Oct. 01).}
\label{image__limit_V1334}
\end{figure}

\section{Detection limit of our binary Cepheids}
\label{section__aplication}

In this section, we set the detection limit for the sample of Cepheids previously presented. For V1334~Cyg, AX~Cir, RT~Aur, and AW~Per, where a companion is detected, we  removed analytically the companion first. Although both methods are estimated by \texttt{CANDID}, all of the given detection limits are derived from our injection method. We present two detection limits for each star, one using all of the observables, and one using only the closure phases. We listed three different values, the average for $r < 50$\,mas, $r < 25$\,mas and $r >25$\,mas, which can be relevant when the limit increases with $r$. All of the final detection limits are listed in Table~\ref{table__limits}, where the values are conservative as they correspond to the mean plus the standard deviation for the given radius range.

From an evolutionary timescale point of view, most of the companions should be stars close to the main sequence. We therefore set upper limits for the spectral type of the companion assuming it is on the main sequence, based on their $H$-band luminosities.

\begin{table*}[]
\centering
\caption{$3\sigma$ average detection limits of the flux ratio.}
\begin{tabular}{c|c|c|c|c|c|c} 
\hline
\hline
                                                                                        &       \multicolumn{2}{c|}{All Observables}    &       \multicolumn{2}{c|}{Only $CP$}                                                                  &       Sp. Type    & \multirow{3}{*}{Instrument}\\
                                                                                        &       \multicolumn{2}{c|}{$r < 50$\,mas}             &       \multicolumn{2}{c|}{$r < 50$\,mas}                                                      &       upper limit           &       \\
                                                                                        &       $r < 25$\,mas      & $r > 25$\,mas         &       $r < 25$\,mas    &  $r > 25$\,mas                                                                &                                                       &       \\
\hline
V1334~Cyg                                               &       \multicolumn{2}{c|}{1.71\,\%}           & \multicolumn{2}{c|}{1.54\,\%}           &               \multirow{2}{*}{B7V}                                    &       \multirow{2}{*}{MIRC}\\
(2012 Jul. 27)                                  &       1.45\,\%        &       1.72\,\%        & 1.19\,\%        &       1.57\,\% &                                      &               \\
\hline
V1334~Cyg                                       &       \multicolumn{2}{c|}{2.59\,\%}           & \multicolumn{2}{c|}{2.44\,\%}                           &                       \multirow{2}{*}{--}             &       \multirow{2}{*}{--}     \\
(2012 Oct. 01)                                  &       2.15\,\%        &       2.60\,\%                        & 1.86\,\%        &       2.42\,\% &                      &                       \\
\hline
AX~Cir                                                          &       \multicolumn{2}{c|}{0.36\,\%}           & \multicolumn{2}{c|}{0.73\,\%}                           &               \multirow{2}{*}{A5V}            &       \multirow{2}{*}{PIONIER} \\
(2013 Jul. 14)                                  &       0.40\,\%        &       0.34\,\%        & 0.80\,\%        &       0.69\,\%        &                       &                       \\
\hline
RT~Aur                                                  &       \multicolumn{2}{c|}{0.64\,\%}           & \multicolumn{2}{c|}{0.47\,\%}           &                       \multirow{2}{*}{F0V}                    &       \multirow{2}{*}{MIRC}   \\
                                                                                &       0.44\,\%        &       0.64\,\%        & 0.33\,\%        &       0.47\,\%        &                       &                       \\
\hline
AW~Per                                                  &       \multicolumn{2}{c|}{0.62\,\%}           & \multicolumn{2}{c|}{0.72\,\%}           &                       \multirow{2}{*}{B9V}                    & \multirow{2}{*}{MIRC}   \\
                                                                                &       0.54\,\%        &       0.62\,\%        & 0.52\,\%        &       0.73\,\%        &                       &                       \\
\hline
SU~Cas                                                  &       \multicolumn{2}{c|}{1.43\,\%}           & \multicolumn{2}{c|}{1.37\,\%}           &                       \multirow{2}{*}{A0V}                            &       \multirow{2}{*}{MIRC} \\
                                                                                &       1.06\,\%        &       1.44\,\%        & 0.99\,\%        &       1.37\,\%        &                       &                       \\
\hline
T~Vul                                                   &       \multicolumn{2}{c|}{1.04\,\%}           & \multicolumn{2}{c|}{1.13\,\%}           &                       \multirow{2}{*}{B9V}                            & \multirow{2}{*}{MIRC}\\
(2012 Jul. 26)                          &       0.84\,\%        &       1.05\,\%        & 0.81\,\%        &       1.14\,\%        &                       &                       \\
\hline
T~Vul                                           &       \multicolumn{2}{c|}{1.21\,\%}           & \multicolumn{2}{c|}{1.29\,\%}           &                       \multirow{2}{*}{--}                     &       \multirow{2}{*}{--}     \\
(2012 Sep. 30)                  &       0.89\,\%        &       1.21\,\%        & 0.92\,\%        &       1.30\,\%        &                       &                       \\
\hline
\end{tabular}
\label{table__limits}
\end{table*}

\paragraph{V1334~Cyg:} This binary system contains a visual and a spectroscopic component. While the visual companion is $> 150$\,mas, the close component was spatially resolved using interferometry by \citet{Gallenne_2013_04_0}, and has a flux ratio of $\sim 3.1$\,\% in $H$ \citep[more details on this companion are presented in][]{Gallenne_2013_04_0}. There is no evidence of a third component so far. After removing  the close companion analytically, we estimated the dynamic range as explained in the previous section. Between the two epochs, we reached a maximum average sensitivity limit of $f_{3\sigma} = 1.54$\,\%. The limit is lower for the first epoch because the atmospheric conditions were better than the second epoch. The contrast upper limit at $3\sigma$ is shown in Fig.~\ref{image__limit_V1334}, with the average values for the two epochs in Table~\ref{table__limits}. This converts to magnitude difference of $\Delta m_\mathrm{H} > -2.5\log{f_{3\sigma}} = 4.5$\,mag. Using the distance $d = 683$\,pc from the $K$-band P-L relation for first overtone (FO) pulsators \citep{Bono_2002_07_0} and the average magnitudes of the Cepheid $H = 4.66$\,mag \citep{Cutri_2003_03_0}, we can exclude the presence of additional companions with a spectral type earlier than B7V stars.

\paragraph{AX~Cir:} The spectroscopic companion was first detected by \citet{Lloyd-Evans_1982_06_0}, and was spatially resolved for the first time with LBI by \citet[][with a more detailed discussion about this companion]{Gallenne_2014_01_0}. The average detection limits are listed in Table~\ref{table__limits}. We reached a dynamic range of $f_{3\sigma} = 0.36$\,\%, corresponding to $\Delta m_\mathrm{H} > 6.1$\,mag if another companion is present. With $d = 500$\,pc \citep[from the $K$-band P-L relation of ][]{Storm_2011_10_0} and $H = 4.66$\,mag for the Cepheid \citep{Cutri_2003_03_0}, we can rule out any other component with a spectral type earlier than A5V star.

\paragraph{RT~Aur:} The binary nature of this short-period Cepheid is still uncertain, however we might have detected it for the first time. Some authors suggested the presence of an early-type companion \citep{Janot-Pacheco_1976_07_0,Balona_1977_01_0}, while others did not see evidence of an additional component  \citep{Harris_1981_05_0,Gieren_1985_07_0}. \citet{Leonard_1986_10_0} summarized various studies from that time about the possible companion orbiting RT~Aur. Based on spectra from the International Ultraviolet Explorer satellite (IUE), \citet{Evans_1992_01_0} did not report any detection and showed that any main-sequence secondary has to be cooler than an A4 star. Radial velocity measurements do not show any variations from orbital motion. Recently, \citet{Turner_2007_11_0} studied long-term photometric light curves and reported a sinusoidal trend consistent with a light travel time effect in the binary system. The companion we might have resolved has a flux ratio of $0.21 \pm 0.12$\,\%, i.e., $\Delta m_\mathrm{H} = 6.7 \pm 0.6$\,mag, leading to $m_\mathrm{H} = 10.6 \pm 1.0$\,mag \citep[using for the Cepheid at our given phase $m_\mathrm{H} = 3.94 \pm 0.01$\,mag from][]{Monson_2011_03_0}. Using the $K$-band P-L relation to get $d = 428$\,pc \citep{Storm_2011_10_0}, we estimate its spectral type to be later than an F1 star. This is compatible with the A4V star upper limit determined by \citet{Evans_1992_01_0} from the International Ultraviolet Explorer (IUE) spectra. However, we need more observations to confirm the existence of the companion as we are at the sensitivity limit of this data set.

Removing this possible companion from our interferometric data, we estimated the detection limit. We reached a minimum dynamic range of $0.47$\,\%, corresponding to $\Delta m_\mathrm{H} > 5.8$\,mag. This is a 90\% completeness azimuthal value, which is a conservative value, and the component at a given position with a slightly higher contrast might be detected (as in this case). This limit allows us to exclude the presence of an additional companion with a spectral type earlier than F0V.

\paragraph{AW~Per:} This Cepheid is a spectroscopic binary with an orbital period of $\sim 40$\,yr. First discovered by \citet{Miller_1964_02_0}, it took several years to derive the first orbit from radial velocity data \citep{Evans_1989_06_0}. It is likely that this companion is itself a binary, as argued by \citet{Evans_1989_06_0,Evans_2000_07_0}, because the magnitude difference between the Cepheid and its companion is not consistent with equal masses and predictions from evolutionary tracks.  Unfortunately, we do not have enough angular resolution and sensitivity to detect this third companion. The properties of the brightest companion were studied based on IUE spectra by \citet{Evans_1994_11_0, Evans_1995_05_0}, who found its spectral type to be B8.3V. This is in agreement with our detection with a flux ratio $f = 1.22 \pm 0.30$\,\%, i.e., a spectral type in the range B6-B9V (using $d = 853$\,pc from a P-L relation and $K = 4.63$\,mag for the Cepheid). We estimated for the companion $m_\mathrm{H} = 9.6 \pm 0.3$\,mag \citep[using for the Cepheid at our given phase $m_\mathrm{H} = 4.84 \pm 0.01$\,mag from][]{Monson_2011_03_0}. \citet{Massa_2008_01_0} also determined the angular separation of the component for another epoch, which  allows us later, with more astrometric points from interferometry, to estimate all the orbital elements, including the inclination and the semi-major axis \citep[see also][]{Gallenne_2013_02_0}.

We derived a maximum sensitivity limit at $3\sigma$ of $0.62$\,\%, i.e., $\Delta m_\mathrm{H} > 5.5$\,mag. We can therefore exclude any additional components with a spectral type earlier than B9V.

\paragraph{SU~Cas:} The binary nature of this Cepheid is still ambiguous. A component was first detected by \citet{Evans_1985_06_0} based on studying the CAII H and K lines, and then a spectral type of B9.5V was determined from IUE spectra \citep{Evans_1991_05_0}. Although the location of SU Cas in a two color diagram is consistent with the presence of a companion, the radial velocity measurements do not show convincing evidence. \citet{Szabados_1991_01_0} found four possible orbital periods from the observations available at that time. Later, \citep{Gorynya_1996_01_0} compiled more data and suggested SU Cas as a possible spectroscopic binary. They derived an orbital period of 408\,days with an eccentricity of $e = 0.43$, but \citet{Groenewegen_2008_09_0} could not confirm the eccentricity with a larger data set and found a period of 407\,days, fixing $e = 0$. Recently, \citet{Remage-Evans_2013_10_0} reanalyzed special dates of Gorynya data, where the velocity difference was supposed to be the largest from the derived orbit, but they concluded that the orbital motion from radial velocity data do not show any convincing detection.

Using the distance $d = 392$\,pc \citep[from the $K$-band P-L relation of first overtone pulsators of][]{Bono_2002_07_0} and a Cepheid magnitude $H = 4.27$\,mag \citep{Cutri_2003_03_0}, a B9.5V companion should give a flux ratio $\sim 1.8$\,\% in the $H$ band, which should be detectable with MIRC. Taking the velocity amplitude of $K_1 = 1\,\mathrm{km\,s^{-1}}$ from \citet[][while \citet{Gorynya_1996_01_0} derived a velocity amplitude of $3\,\mathrm{km\,s^{-1}}$]{Groenewegen_2008_09_0}, we estimated $a\,\sin{i} > 0.1$\,mas. Our nondetection could be explained if the companion was located at $< 0.5$\,mas because it would not have been spatially resolved, but this type of close component would have an effect on the radial velocities (unless the orbit is face-on, but the probability of this kind of an orbit is low). The other possibility would be that this companion has a wider orbit. We therefore searched within a 100\,mas radius range, but we did not find a significant detection. We then estimated the sensitivity limit for $r < 100$\,mas, and found a maximum $3\sigma$ flux ratio of 1.65\,\%. This means that if the B9.5V companion was within 100\,mas, we would have detected it. Therefore, if this companion exists, it should have a wider orbit. 

The average sensitivity limits for a given radius range are tabulated in Table~\ref{table__limits}. We reached a mean contrast of 1.37\,\%, which means that $\Delta m_\mathrm{H} > 4.6$\,mag. This converts to an upper limit of the spectral type of an A0V star.

\paragraph{T~Vul:} As with SU~Cus, the radial velocities of this Cepheid do not show any signature of an orbiting companion, while a hot A0.8V component was detected by \citet{Evans_1992_07_0}. In the literature, we found contradictory estimates of the orbital period from radial velocity measurements; for instance, \citet{Kovacs_1990_03_0} found a long period modulation of 738\,d, \citet{Szabados_1991_01_0} estimated $P = 1745$\,d from a larger data set, while \citet{Bersier_1994_11_0} did not find any orbital motion larger 0.55\,km\,s$^{-1}$ using additional more accurate data. They showed that the long period of \citet{Szabados_1991_01_0} is not compatible and argued that the 738\,d period might be an artifact of the time sampling because the observations were only made  in autumn. \citet{Kiss_2000_05_0} reached the same conclusion with additional measurements showing no signature of orbital motion in the radial velocity curve.

From our interferometric observations, we did not detect any companion within 50\,mas. The A0.8V component detected from IUE spectra would correspond to a flux ratio in $H$ of $\sim 0.7$\,\%. According to our estimated interferometric detection limit, listed in Table~\ref{table__limits}, this  kind of a component is below our detection level. We reached an average sensitivity limit of 1.04\,\%, and we can therefore exclude any other possible companion with a spectral type earlier than a B9V star.

\section{Conclusion}
\label{section__conclusion}

We presented an overview of \texttt{CANDID}, a new tool to search for point-source companions and estimate the sensitivity level from interferometric observations using the squared visibilities, closure phases, and amplitude of the bispectrum, when available. \texttt{CANDID} allows us to:

\begin{itemize}
\item efficiently detect companions using a grid of fit and determine the detection level by giving the number of sigmas;
\item set the detection limit for a companion in data where a companion has not been detected; and
\item set the detection limit for a tertiary companion, in the case where a companion has been detected.
\end{itemize}

We used \texttt{CANDID} to investigate a sample of binary Cepheids. We first determined the detection level for our previous detections \citep{Gallenne_2013_04_0,Gallenne_2014_01_0} and showed that the components were detected at $> 13\sigma$ for AX~Cir and $> 25\sigma$ for V1334~Cyg. We also reported a new detection for AW~Per, with a detection level $> 15\sigma$; the companion is located at $\rho = 32$\,mas and $PA = 67^\circ$, with a flux ratio of $f = 1.22$\,\%. The companion orbiting RT~Aur might have been detected at $3.8\sigma$, using only the closure phase signal, however, more observations are needed to confirm the presence of this component. Any additional companions were not detected signficantly (i.e., with a detection level $> 3\sigma$) around these stars.  Likewise, no companions were detected around SU~Cas and T~Vul. From these interferometric data, we were able to set upper limits for the spectral types; we found no components with a spectral type earlier than B7V, A5V, F0V, B9V, A0V, and B9V for V1334~Cyg, AX~Cir, RT~Aur, AW~Per, SU~Cas, and T~Vul, respectively.

The fitting procedure of \texttt{CANDID} also allowed us to measure the uniform disk angular diameters of the new Cepheids observed.  We found for RT~Aur $\theta_\mathrm{UD} = 0.699 \pm 0.011$\,mas (at the pulsation phase $\phi = 0.32$), for AW~Per $\theta_\mathrm{UD} = 0.627 \pm 0.018$\,mas (at $\phi = 0.52$), for SU~Cas $\theta_\mathrm{UD} = 0.609 \pm 0.043$\,mas (at $\phi = 0.77$), and $\theta_\mathrm{UD} = 0.608 \pm 0.013$\,mas and $\theta_\mathrm{UD} = 0.635 \pm 0.018$\,mas for T~Vul (at $\phi = 0.27$ and $\phi = 0.12$).

We demonstrated that the approximation we used to analytically inject a companion and estimate the detection limits is valid (i.e., error $< 0.5$\,\%) for contrasts $f \leqslant 5$\,\% if we use all of the observables, and up to $f \leqslant 50$\,\% using only the squared visibilities and the closure phases. This makes \texttt{CANDID} a useful tool for analyzing long-baseline interferometric observations of binary star systems.

Finally, this work demonstrates the capabilities of the MIRC and PIONIER instruments, which can reach a dynamic range of 1:200, depending on the angular distance of the companion and the ($u,v$) plane coverage. In the future, we plan to work on improving the sensitivity limits for realistic data through better handling of the correlations.


\begin{acknowledgements}
The authors would like to thank the CHARA Array and Mount Wilson Observatory staff for their support. Research conducted at the CHARA Array is funded by the National Science Foundation through NSF grant AST-0908253, by Georgia State University, the W. M. Keck Foundation, the Packard Foundation, and the NASA Exoplanet Science Institute. The authors also thank all the people involved in the VLTI project. A.~G. acknowledges support from FONDECYT grant 3130361. JDM acknowledges funding from the NSF grants AST-0707927, AST-0807577, and AST 1108963. We also acknowledge the STSCI grant HST-GO13841.006-A and HST-GO-13454.07-A. WG and GP gratefully acknowledge financial support for this work from the BASAL Centro de Astrof\'isica y Tecnolog\'ias Afines (CATA) PFB-06/2007. Support from the Polish National Science Centre grant MAESTRO DEC-2012/06/A/ST9/00269 and the Polish Ministry of Science grant Ideas Plus (awarded to G.~P.) is also acknowledge. W.~G. also acknowledges financial support from the Millenium Institute of Astrophysics (MAS) of the Iniciativa Cientifica Milenio del Ministerio de Econom\'ia, Fomento y Turismo de Chile, project IC120009. A.~G. also acknowledge computational resources from FONDECYT grant 1130521 and 1130721. We also acknowledge support from the ECOS/Conicyt grant C13U01. PIONIER is funded by the Universit\'e Joseph Fourier (UJF), the Institut de Plan\'etologie et d'Astrophysique de Grenoble (IPAG), the Agence Nationale pour la Recherche (ANR-06-BLAN-0421, ANR-10-BLAN-0505, ANR-10-LABX56), and the Institut National des Science de l'Univers (INSU PNP and PNPS). The integrated optics beam combiner is the result of a collaboration between IPAG and CEA-LETI based on CNES R\&T funding. This research received the support of PHASE, the high angular resolution partnership between ONERA, Observatoire de Paris, CNRS, and University Denis Diderot Paris 7. This work made use of the SIMBAD and VIZIER astrophysical database from CDS, Strasbourg, France and the bibliographic informations from the NASA Astrophysics Data System. This research has made use of the Jean-Marie Mariotti Center \texttt{SearchCal} and \texttt{ASPRO} services, co-developed by FIZEAU and LAOG/IPAG, and of CDS Astronomical Databases SIMBAD and VIZIER.
\end{acknowledgements}


\bibliographystyle{aa}   
\bibliography{/Users/alex/Sciences/Articles/bibliographie}

\end{document}